\documentclass[twocolumn,groupeaddress,showpacs]{revtex4}
\usepackage{psfig}
\usepackage{epsfig}

\begin{document}
\title{Capacitively coupled hot-electron nanobolometer as far-infrared photon counter}
\date{\today}

\author{Drago\c s-Victor Anghel$^{\rm a)}$\footnotetext{$^{\rm a)}$E-mail: 
{\tt dragos@fys.uio.no}}} 
\affiliation{University of Oslo, Department of Physics, P.O.Box 1048 
Blindern, N-0316 Oslo, Norway}
\affiliation{NIPNE  -- ``HH'', P.O.Box MG-6, 
R.O.-76900 Bucure\c sti - M\u agurele, Romania}
\author{Leonid Kuzmin$^{\rm b)}$\footnotetext{$^{\rm b)}$E-mail: 
{\tt kuzmin@fy.chalmers.se}}}
\affiliation{Department of Microelectronics and Nanoscience, Physics 
and Engineering Physics, Chalmers University of Technology, S-41296 
G\"oteborg, Sweden}

\begin{abstract} 
We show theoretically that hot-electron nanobolometers 
consisting of a small piece of normal metal, capacitively 
coupled to a superconducting antenna through a pair of normal 
metal--insulator--superconductor (NIS) tunnel junctions may be used as 
far-infrared photon counters. To make the device 
most effective at high counting rates, we suggest the use of the 
bolometer in the simplest configuration, when the NIS tunnel junctions are 
used as both an electron cooler and thermometer. 
The absorption of the photon in the normal metal 
produces a pulse in the electron temperature, which is measured by the 
NIS junctions. The counter may resolve photons up to 0.3--0.4 mm 
wavelength and has a typical re-equilibration time constant of about 20 ns.
\end{abstract}

\maketitle

The far-infrared region is one of the richest areas of spectroscopic 
research, with applications ranging from molecular physics to astronomy. 
These applications, especially the astrophysical observations, require
extremely sensitive detectors. 
Depending on the energy of the incident photons, the input 
power, and the response time of the detector, the device may resolve 
individual quanta (quantum detectors), or may measure the total input 
power (integrating detectors). 
In general, integrating detectors measure long-wavelength 
radiation fluxes (infrared and far infrared regions 
\cite{integrating,NHEB-CC}), while high energy photons (for example, 
x ray) are observed by quantum detectors. 
The first x-ray detector using a normal metal film as an absorber and a 
normal metal--insulator--superconductor (NIS) tunnel junction as a 
thermometer, was built by Nahum {\em et al.} \cite{X-ray}.
The photon interacts with the detector by producing a {\em hot spot} in 
the normal metal. The energy dissipated into the hot spot diffuses then into 
the whole absorber and the increase of electron temperature is measured by 
the NIS thermometer. The re-equilibration time constant of the detector 
was of the order of 10 ${\rm \mu s}$ 
and determined mostly by the electron--phonon coupling.
The same phenomenon of hot spot formation due to the photon absorption 
was suggested for the detection of longer 
wavelength photons \cite{hotspot}. Recently, Semenov 
et al. \cite{semenov} constructed a near-infrared photon counter based on a 
current biased narrow superconducting strip. The hot spot formed 
in the strip leading to a break of the superconductivity across it, 
which further produced a signal in the voltage read out. At the 
experimental stage reported until now, the time constant of the detector 
is of the order of 100 ps and is limited by the amplifier 
bandwidth--the estimated thermalization time 
of the superconducting strip being of the order of tens of picoseconds.

\begin{figure}[t]
\begin{center}
\unitlength1mm\begin{picture}(80,45)(0,0)
\put(0,0){\epsfig{file=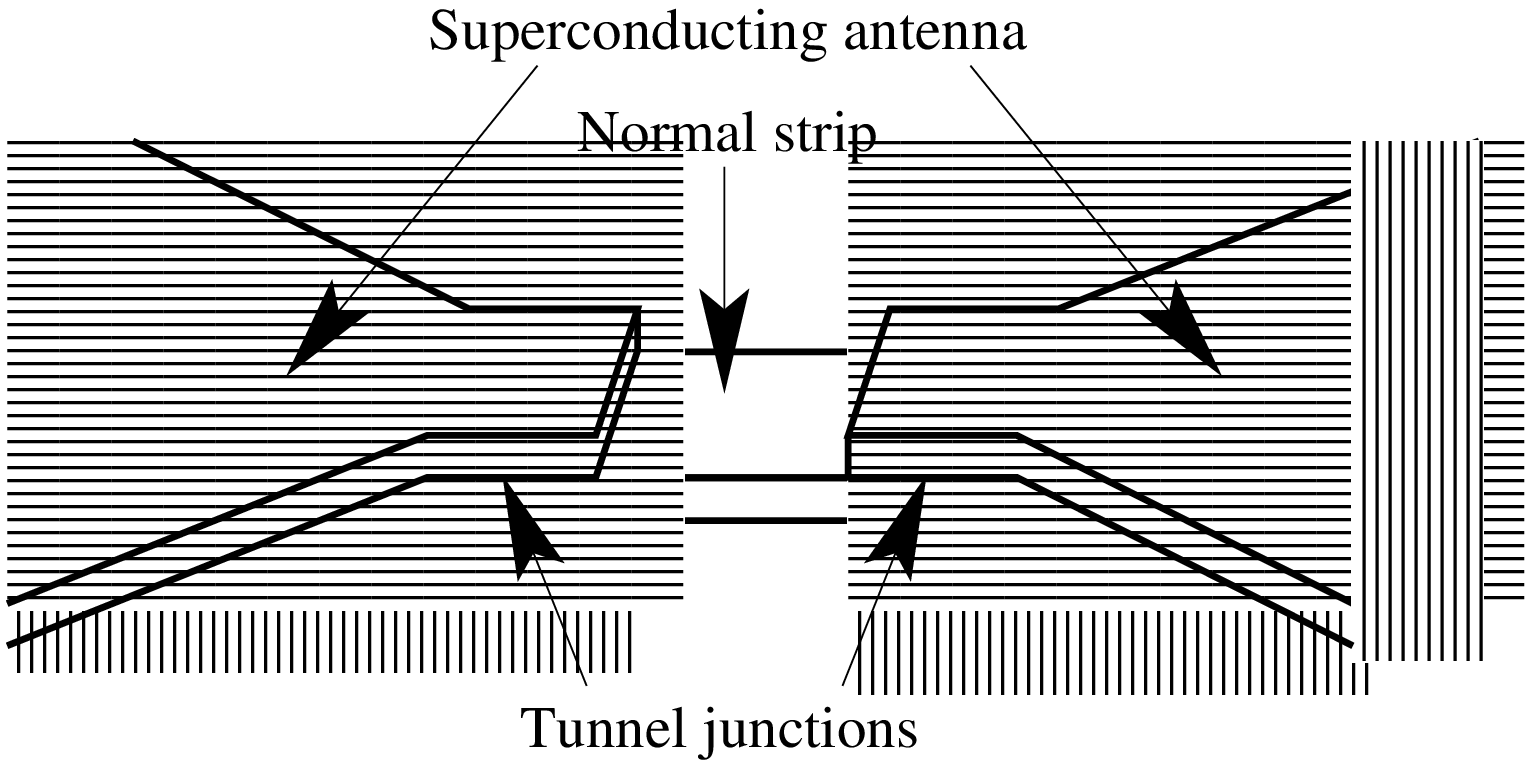,width=80mm}}
\end{picture}
\caption{Schematic drawing of the main part 
of the photon quantum detector.}
\label{antenna}
\end{center}
\end{figure}

To increase the limit wavelength of detectable individual 
IR photons, at a counting rate high enough for astronomical observations 
but still within the range of broad-bandwidth amplifiers, we suggest the 
use of the capacitively coupled hot-electron microbolometer \cite{NHEB-CC} 
as a photon counter. 
The detector (Fig. \ref{antenna}) is formed of a small normal metal 
island connected to the superconducting antenna by two symmetric NIS
tunnel junctions.
As in the case of x-ray detectors \cite{X-ray}, the energy of the quanta, 
dissipated in the normal metal island [which we call the {\em thermal 
sensing element} (TSE)] leads to an increase of the electron 
temperature, which is measured by  the NIS tunnel junctions.
The time scale of the heat diffusion into the whole normal metal 
element is $\tau_{\rm d} \sim L^2/(\pi^2 D)$, where $L$ is the 
linear dimension of the sample and $D$ is the electron diffusion 
constant. If we set $L=1 \ {\rm \mu m}$, then, for a typical value 
of $D=10^{-4} \ {\rm m^2s^{-1}}$ in Cu thin films, 
$\tau_{\rm d} \approx 1$ ns. 
To have an accurate reading of the temperature variation in the TSE, 
the detector re-equilibration time, $\tau$, 
should be larger than $\tau_{\rm d}$. 
Although not shown in Fig. \ref{antenna}, the detector 
(TSE plus antenna) is supported by an insulating substrate, 
which acts also as a heat bath of temperature $T_{\rm b}$. 
Here, we assume perfect thermal contact between the substrate and the 
detector lattice, and we set the lattice temperatures of both the TSE and 
the antenna equal to $T_{\rm b}$. 
On the other hand, in the operating temperature range of the detector
(of the order of 100 mK) the electron--phonon coupling is weak and 
we assign to the electron gas in the TSE an effective 
temperature $T_{\rm e}$, which, in general, is different from $T_{\rm b}$. 
For the most effective operation of the nanobolometer, 
the NIS junctions are used both to cool the electron gas in the 
detector bellow the heat bath temperature $T_{\rm b}$ and to measure 
the temperature pulse (the principles 
of NIS cooling and temperature measurement are presented in 
Refs. \cite{NIScooler,jukka_c} and \cite{nahum_bolo}, respectively). 
In the normal metal, the power 
flow from the electron system to the lattice is 
$\dot{Q}_{\rm ep} = \Sigma_{\rm ep} \Omega (T_{\rm e}^5 - T_{\rm b}^5)$ 
\cite{wellstood}, where $\Omega$ is the 
volume of the TSE. 
We approximate the electron specific heat with 
$C_{\rm V}=\Omega(k_{\rm B}^2T_{\rm e}/3\hbar^3)\sqrt{2 m_{\rm e}^3 \epsilon_{\rm F}}$, where $\epsilon_{\rm F}$ 
is the Fermi energy and $m_{\rm e}$ is the electron mass. 
In what follows, for the concrete calculations, we shall consider Cu as 
the normal metal and Al as the superconductor since these two metals are 
routinely used in the construction of microrefrigerators.
The cooling power of one junction is denoted by $\dot{Q}_{\rm J}$ and, 
for a symmetric setup as the one shown in Fig. \ref{antenna}, the 
total power extracted through the junctions is $2\dot{Q}_{\rm J}$. 
$\dot{Q}_{\rm J}$ is a function of both the effective temperature of the 
quasiparticles in the superconductor, $T_{\rm s}$, and $T_{\rm e}$. 
We assume that the superconductor is thermalized by using 
efficient traps \cite{noitrap}, so we use also $T_{\rm s}=T_{\rm b}$. 
%
At equilibrium (no incident radiation), the total power extracted from 
the normal metal is zero: 
\begin{equation} \label{balance}
\dot{Q}_{\rm T} \equiv 2 \dot{Q}_{\rm J} + \dot{Q}_{\rm ep} = 0 \, .
\end{equation}
Equation (\ref{balance}) sets the equilibrium electron temperature, 
$T_{\rm e0}$, as a function of $T_{\rm b}$.
As a thermometer, the NIS junctions may work in either voltage biased 
(VB) or current biased (CB) regime. 
For a linear response of the detector, the temperature pulse due to 
the photon absorption, 
$\delta T_{\rm e}\equiv T_{\rm e} - T_{\rm e0} \ll T_{\rm e}$, 
has an exponential decay 
$\delta T_{\rm e}(t) = \delta T_{\rm e}(0)\exp{(-t/\tau)}$, 
where $\delta T_{\rm e}(0)=\hbar\omega/C_{\rm V}$ and we assume that 
$\tau\equiv C_{\rm V}\left(\partial\dot{Q}_{\rm T}/\partial T_{\rm e}\right)_{T_{\rm e}=T_{\rm e0}}^{-1} \gg \tau_{\rm d}$. 
Using Eq. (\ref{balance}) and the expression for $\tau$, we write 
$\tau^{-1} = \tau_{\rm J}^{-1} + \tau_{\rm ep}^{-1}$, in obvious 
notations. For Cu at $T_{\rm e}=100$ mK 
($\Sigma_{\rm ep} = 4 \ {\rm nW K^{-5}\mu m^{-3}}$ \cite{jukka_c}), 
$\tau_{\rm ep} \approx 3.5 \times 10^{-5}$ s, which will turn out to be 
much larger than $\tau_{\rm J}$. 

Besides $\tau$, the most important figure of merit of the detector is the 
energy resolution (ER). A photon can be detected if the signal 
produced by its absorption is larger than the square root of the mean square 
fluctuation of the measured quantity, 
$\langle \delta^2 M\rangle^{1/2}\equiv\langle (M-\langle M\rangle)^2\rangle^{1/2}$, where $M$ is current (VB) or voltage (CB), and by 
$\langle\cdot\rangle$ we denote the statistical average. 
The ER can then be defined as $\langle\delta^2M\rangle^{1/2}/m_{\rm max}$, 
where $m_{\rm max}$ is the amplitude of the {\em measured} pulse. 
Due to the finite bandwidth of the measuring device, $\tau_{\rm c}^{-1}$, 
the measured and real values of $M$, say $M_{\rm m}(t)$ and $M_{\rm r}(t)$, 
respectively, may be related in a general formalism by the equation 
$dM_{\rm m}(t)=\tau_{\rm c}^{-1}[M_{\rm r}(t) - M_{\rm m}(t)]dt$.
This leads to the usual frequency dependent  amplification 
factor, proportional to $[1+ (\omega \tau_{\rm c})^2]^{-1/2}$. 
Therefore, the {\em measured} mean square fluctuation is
%
$\langle \delta^2 M_{\rm m}\rangle=\int_0^\infty\langle\delta^2 M_{\rm r}\rangle_\omega/[1+ (\omega \tau_{\rm c})^{2}] d\omega$, 
where $\langle\delta^2 M_{\rm r}\rangle_\omega$ is the spectral density 
of noise. 
In VB and CB regimes, $M_{\rm r}\equiv I_{\rm J}$ (current 
through the junctions) and $M_{\rm r}\equiv 2V$ (voltage across both 
junctions), respectively. 
Following Ref. \cite{euj1} and disregarding the external circuit of 
the detector, we write the total fluctuation $\delta I_{\rm J}$ (VB) 
as the superposition of 
the fluctuation due to the discrete transport of 
charges through the junctions, and the current fluctuations induced by 
the temperature and particle fluctuations in the normal metal island. 
In the CB regime, the voltage fluctuation is induced only by the 
particle number fluctuation. 
Using this, we obtain the following spectral densities: 
\begin{eqnarray}
\langle \delta^2 I_{\rm J} \rangle_\omega &=& 
\langle \delta^2 I_{\rm J,shot} \rangle_\omega +
\left( \frac{\partial I_{\rm J}}{\partial T_{\rm e}} \right)^2
\langle\delta^2 T_{\rm e}\rangle_\omega \nonumber \\
&& + \frac{1}{e^2\omega^2}\left(\frac{\partial\epsilon_{\rm F}}{\partial N}
\frac{\partial I_{\rm J}}{\partial (eV)}\right)^2
\langle\delta^2I_{\rm J,shot}\rangle_\omega \nonumber \\
& & + \left.
2\frac{\partial I_{\rm J}}{\partial T_{\rm e}}  
Re \Big(\langle \delta I_{\rm J,shot} \delta T_{\rm e} \rangle_\omega \Big)
\right. \nonumber\\ 
& & + 2 \frac{\partial\epsilon_{\rm F}}{\partial N}
\frac{\partial I_{\rm J}}{\partial (eV)} 
\frac{\partial I_{\rm J}}{\partial T_{\rm e}}
Re \left(\left\langle \delta T_{\rm e} \frac{\delta \dot{N}}{i\omega} 
\right\rangle_\omega \right) \, ,
\label{fluctI} \\
\langle \delta^2 V \rangle_\omega &=& \frac{1}{\omega^2} 
\left(\frac{1}{e^2}\frac{\partial\epsilon_{\rm F}}{\partial N}\right)^2 
\langle \delta^2 I_{\rm J,shot} \rangle_\omega ,
\label{fluctV} 
\end{eqnarray}
where $\epsilon_{\rm F}$ is the Fermi energy, $N$ is the number of 
electrons in the normal metal, and 
$\langle\delta^2 N\rangle_\omega=\langle \delta^2 I_{\rm J} \rangle_\omega\cdot(\omega e)^{-2}$. We also assumed that 
$\partial (eV)/\partial N\equiv\partial\epsilon_{\rm F}/\partial N=(2/3)(\epsilon_{\rm F}/N)$, as for the ideal gas. The method to evaluate 
$\langle\delta^2 T_{\rm e}\rangle_\omega$ and the correlation terms was 
given in Ref. \cite{euj1}. 
In each case, the amplifier noise should be added quadratically to the 
fluctuations above. $\langle\delta^2I_{\rm J,shot}\rangle_\omega$ 
is the current Poissonian shot noise (white noise). 
According to Eqs. (\ref{fluctI}) and (\ref{fluctV}), in both VB and CB 
regimes $\langle\delta^2 M_{\rm r}\rangle_\omega\propto \omega^{-2}$ for 
$\omega\to 0$, which implies an ``infrared'' divergence of the total measured 
fluctuation $\langle \delta^2 M_{\rm m}\rangle$. 
Apparently, this fact would make our approach totally hopeless. 
Fortunately we can avoid this, at least for the VB regime. 
From Eq. (\ref{fluctI}) we notice that 
$\langle \delta^2 I_{\rm J} \rangle_\omega$ is the sum of an 
$\omega$-independent term, call it 
$\langle \delta^2 I_{\rm J0} \rangle$, and an $\omega$-dependent one, 
say $\langle \delta^2 I_{\rm Jd} \rangle_\omega$. Obviously, 
$\langle\delta^2 I_{\rm Jd}\rangle_\omega\to 0$ or $\infty$ as 
$\omega\to\infty$ or $\omega\to 0$, respectively. 
We can define the crossover 
between $\omega$-independent and $\omega$-dependent regimes as 
$\langle\delta^2I_{\rm J}\rangle_{\omega=\omega_{\rm c}}^{1/2}\cdot\langle\delta^2I_{\rm J}\rangle_{\omega\to\infty}^{-1/2}=2$. 
If $\omega_{\rm c}\ll 1/\tau$ and since ``slow'' fluctuations 
of $I_{\rm J}$ do not influence the reading of the signal due to the 
photon absorption, we can approximate the total fluctuation 
by using a convenient cutoff of the integral at the lower end:
\begin{eqnarray}
\langle \delta^2 I_{\rm Jm}\rangle &\approx& \int_{\omega_0}^\infty
\frac{\langle\delta^2 I_{\rm J}\rangle_\omega}{1+ (\omega \tau_{\rm c})^2} 
d\omega 
\le \frac{\pi}{2}\frac{\langle\delta^2 I_{\rm J0}\rangle}{\tau_{\rm c}}
, \label{I_fluct}
\end{eqnarray}
where $\omega_{\rm c}\ll\omega_0\ll 1/\tau$.
From Eq. (\ref{fluctV}), we see that this procedure cannot be 
applied to CB regime. 
Moreover, evaluations of $\langle \delta^2 I_{\rm Jm} \rangle$ 
and $\langle \delta^2 V_{\rm m} \rangle$, using a similar cutoff show that 
the second regime is not practical, and 
in what follows we shall consider only the VB regime.

Let us focus now on the signal measurement. In the linear regime, 
if the photon is absorbed at $t=0$, then 
$I_{\rm J}(t<0) \equiv I_{\rm J}(T_{\rm e0})$, while 
$I_{\rm J}(t\ge 0)=I_{\rm J}(T_{\rm e0}) + i_0e^{-t/\tau}$ ,
where $i_0\equiv I_{\rm J}[T_{\rm e}(0)]-I_{\rm J}(T_{\rm e0})$. 
If $\tau_{\rm c} \gg \tau_{\rm d}$, the measured current is  
$I_{\rm Jm}(t) = I_{\rm J}(T_{\rm e0})+i_{\rm m}(t)$, where 
$i_{\rm m}(t<0)=0$ and 
$i_{\rm m}(t\ge 0)=i_0\tau(\tau_{\rm c}-\tau)^{-1}\left(e^{-t/\tau_{\rm c}}-e^{-t/\tau}\right)$. 
%
If we denote $x\equiv \tau_{\rm c}/\tau$, then the maximum value of 
$i_{\rm m}(t)$ is 
$i_{\rm max} = i_0 x^{x/(1-x)}$
and provides a way to calculate the energy of the absorbed photon.
According to the definition, the relative error in the determination of $i_0$, 
$\lambda = \langle \delta^2 I_{\rm Jm}\rangle^{1/2}i_{\rm max}^{-1}$,
is also the ER. If $\hbar\omega_{\rm ph}$ 
is the energy of the incoming photon, in the linear approximation 
$i_0\approx(\partial I_{\rm J}/\partial T_{\rm e})_{T_{\rm e0}}[\hbar\omega_{\rm ph}/C_{\rm V}(T_{\rm e0})]\ll I_{\rm J}(T_{\rm e0})$. Plugging this 
together with (\ref{I_fluct}) in the expression for ER, we get
\begin{equation} \label{ER}
{\rm ER} = x^{-\frac{1+x}{2(1-x)}}
\sqrt{\frac{\pi\langle \delta^2 I_{\rm J0}\rangle}{2\tau}}
\frac{C_{\rm V}(T_{\rm e})}{\left( \partial I_{\rm J}/\partial T_{\rm e}
\right)_{T_{\rm e0}} \hbar\omega} .
\end{equation} 
Since $x^{-(1+x)/2(1-x)}$ has a minimum equal to $e$ for $x=1$, 
we obtain the {\em optimum energy resolution}, 
${\rm ER_{opt}}=e\left[(\pi/2)\langle\delta^2 I_{\rm J0}\rangle/\tau \right]^{1/2} C_{\rm V}(T_{\rm e})\left[\left(\partial I_{\rm J}/\partial T_{\rm e}\right)\cdot\hbar\omega\right]^{-1}$. 

\begin{figure}[t]
\begin{center}
\unitlength1mm\begin{picture}(80,64)(0,0)
\put(0,0){\epsfig{file=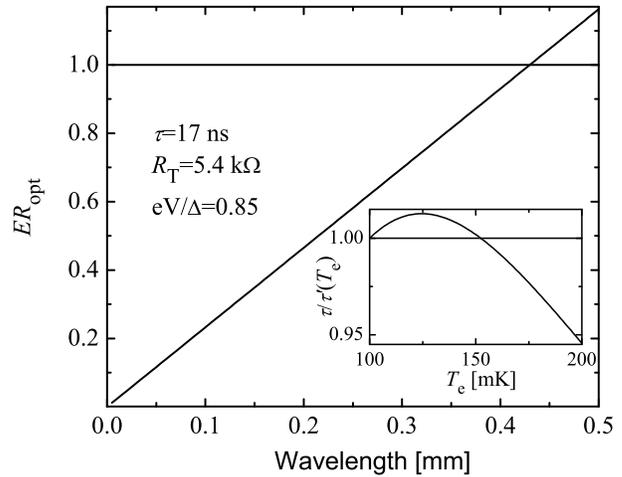,width=80mm}}
\end{picture}
\caption{Energy resolution as a function of the incident photon
wavelength. The inset shows the relative variation of the time 
constant: $\tau/\tau'(T_{\rm e})$. $T_{\rm b}=0.3$ K and 
$T_{\rm e0}=0.1$ K.}
\label{ermax}
\end{center}
\end{figure}

We now turn to concrete calculations. 
For this, we take a typical value: $T_{\rm b}=300$ mK, 
and we set our goal to $T_{\rm e0}=100$ mK. 
If we set $\Omega=2\times 10^{-3}\ {\rm \mu m}^3$, then 
$T_{\rm e}$ increases to about 200 mK after the absorption of a 
photon of 100 ${\rm \mu m}$ wavelength. 
With all these parameters fixed, Eq. (\ref{balance}) is an equation in the 
junction tunnel resistance, $R_{\rm T}$, and $V$. Using this equation, 
we write ${\rm ER_{opt}}$ as a function of $V$. 
This function has a minimum at $V=V_{\rm opt}\approx 0.85\Delta/e$ 
($\Delta$ is the gap energy in the superconductor). 
Setting the bias voltage at $V_{\rm opt}$, we obtain 
$R_{\rm T}\approx 5.4\ {\rm k\Omega}$ (Eq. \ref{balance}) and 
$\tau\approx 17$ ns. 
Moreover, in this case 
$\tau\cdot\omega_{\rm c}\approx 1/23\ll 1$ which justifies 
the assumption behind Eq. (\ref{I_fluct}) and, therefore, the 
expression for ${\rm ER_{opt}}$. In Fig. \ref{ermax}, we show 
$\left.{\rm ER_{opt}}\right|_{V_{\rm opt},T_{\rm e0}}$ 
(which was calculated using $\langle\delta^2 I\rangle_{\omega=\tau^{-1}}$ 
instead of $\langle\delta^2 I_{\rm J0}\rangle$, to set 
an upper limit for nonlinear effects in ${\rm ER_{opt}}$) as a function 
of the incident photon wavelength. From this plot, it appears that 
the device may work as a counter up to about 0.4 mm photon wavelength. 
For longer wavelengths, 
${\rm ER_{opt}}$ is above 1, so, in principle, the signal may not be 
distinguished from the noise. Nevertheless, the performances may be 
improved by decreasing $T_{\rm b}$ and $T_{\rm e}$. If 
$\delta T_{\rm e}\ll T_{\rm e0}$ is not satisfied, the response 
time would show a 
variation with the temperature. If we define the temperature dependent 
re-equilibration time, 
$\tau'(T_{\rm e})\equiv -(T_{\rm e}-T_{\rm e0})/(dT_{\rm e}/dt)$, 
we observe that this does not vary much in the range of interest (see 
inset of Fig. \ref{ermax}). 
The approximation $\tau'(T_{\rm e})\approx\tau'(T_{\rm e0})=\tau$ 
was used for the convenience 
of performing analytical calculations. For a concrete experimental 
set-up one can calculate numerically the energy resolution, taking into 
account nonlinear effects. According to the 
evaluations above, the results should not differ significantly.

In conclusion, we presented a far-infrared photon counter formed of a 
small piece of normal metal, capacitively coupled to a superconducting 
antenna via two NIS tunnel junctions. 
The photon energy is released in the normal metal and the junctions 
serve as both an electron cooler and thermometer. 
At a bath temperature $T_{\rm b}=300$ mK, the electron temperature 
in the normal metal may be reduced to $T_{\rm e0}=100$ mK by the 
cooling effect of the junctions, which tunes the 
response time of the detector (time constant for the detector 
re-equilibration after one photon absorption) to $\tau\approx 17$ ns and 
the tunnel resistance of each junction to 
$R_{\rm T}\approx 5.4\ {\rm k\Omega}$. 
For a volume of the normal metal $\Omega=2\times 10^{-3}\ {\rm \mu m}^3$, 
the calculated ER is 0.23 at a photon 
wavelength $\lambda=0.1$ mm (Fig. \ref{ermax}). The ER reaches 1 
(the counting limit) 
for $\lambda\approx 0.4$ mm. To allow for counting longer-wavelength 
photons, one should retune the detector parameters. The technological 
and physical 
difficulties raised eventually by the construction of small tunnel 
junctions with such low transparencies may be overcome either by the 
use of ferromagnetic materials to suppress the Andreev current 
across the too thin barriers, as suggested by Giazotto {\em et al.} 
\cite{gigi}, or by the use of heavily doped semiconductors, which 
would allow one to increase the size of the absorber \cite{pruna}.

We thank A. Semenov, H. W. H\"ubers, G. N. Gol'tsman, for fruitful 
discussions. 
This research was supported by Swedish Institute  and Research Council
grants.

\end{document}